\documentclass[apj,numberedappendix]{emulateapj}

\slugcomment{Accepted January 16, 2012} 

\usepackage{graphicx}
\usepackage{hyperref}
\usepackage{bm}
\usepackage{CJK}

\shorttitle{Magnetic Evolution from \textit{SDO}/HMI}
\shortauthors{Sun et al.}

\begin{document}

\title{Evolution of Magnetic Field and Energy in A Major Eruptive Active Region Based on \textit{SDO}/HMI Observation}

\author{Xudong Sun\altaffilmark{1,2}, J. Todd Hoeksema\altaffilmark{1}, Yang Liu\altaffilmark{1}, 
  Thomas Wiegelmann\altaffilmark{3}, Keiji Hayashi\altaffilmark{1}, Qingrong Chen\altaffilmark{2}, Julia Thalmann\altaffilmark{3}}

\email{xudongs@stanford.edu}

\altaffiltext{1}{W. W. Hansen Experimental Physics Laboratory, Stanford University, Stanford, CA 94305, USA.}
\altaffiltext{2}{Department of Physics, Stanford University, Stanford, CA 94305, USA.}
\altaffiltext{3}{Max-Planck-Institut f\"{u}r Sonnensystemforschung, Max-Planck-Str. 2, 37191 Katlenburg-Lindau, Germany.}


\begin{abstract}
We report the evolution of magnetic field and its energy in NOAA active region 11158 over 5 days based on a vector magnetogram series from the Helioseismic and Magnetic Imager (HMI) on board the \textit{Solar Dynamic Observatory} (\textit{SDO}). Fast flux emergence and strong shearing motion led to a quadrupolar sunspot complex that produced several major eruptions, including the first X-class flare of Solar Cycle 24. Extrapolated non-linear force-free coronal fields show substantial electric current and free energy increase during early flux emergence near a low-lying sigmoidal filament with sheared kilogauss field in the filament channel. The computed magnetic free energy reaches a maximum of $\sim$$2.6\times10^{32}\;\rm{erg}$, about $50\%$ of which is stored below $6\;\rm{Mm}$. It decreases by $\sim$$0.3\times10^{32}\;\rm{erg}$ within 1 hour of the X-class flare, which is likely an underestimation of the actual energy loss. During the flare, the photospheric field changed rapidly: horizontal field was enhanced by $28\%$ in the core region, becoming more inclined and more parallel to the polarity inversion line. Such change is consistent with the conjectured coronal field ``implosion'', and is supported by the coronal loop retraction observed by the Atmospheric Imaging Assembly (AIA). The extrapolated field becomes more ``compact'' after the flare, with shorter loops in the core region, probably because of reconnection. The coronal field becomes slightly more sheared in the lowest layer, relaxes faster with height, and is overall less energetic.
\end{abstract}



\section{Introduction}
\label{sec:intro}

Extreme solar activity is powered by magnetic energy \citep[e.g.][]{forbes2000}. Flux emergence and shearing motion introduce strong electric currents and inject energy to the active region (AR) corona. Coronal fields are subsequently reconfigured, accumulating large amounts of magnetic free energy. When the overly energetic field gets destabilized, part of its excess energy is released rapidly, with power enough to drive explosive phenomena such as flares and coronal mass ejections (CMEs). During this process, fast, irreversible changes of the photospheric field have been observed as the possible imprint of the coronal activity, including transverse field and magnetic shear increase \citep[e.g.][]{wanghm1994,wanghm2006} and longitudinal field decrease \citep{sudol2005}. \cite{hudson2000} suggested that the energy loss during the explosion causes ``implosion'' of the coronal field, which is supported by recent observations \citep{ji2007,liur2010}.

This ``storage-release'' picture provides a scenario successful in explaining many observed phenomena \citep[e.g.][]{schrijver2009}. However, detailed knowledge of the process is still lacking, especially in a quantitative spatially and temporally resolved manner \citep[e.g.][]{hudson2011}. In this sense, uninterrupted, frequent photospheric vector field observation may be crucial, as it directly monitors the AR's evolution and provides information on its non-potential nature \citep[e.g.][]{schrijver2005}. In addition, field extrapolation models based on the vector boundary may provide valuable diagnostics of the changing coronal field \citep[e.g.][]{regnier2006,thalmann2008,jing2009}. The recently launched Heliospheric and Magnetic Imager \citep[HMI;][]{schou2011} on board the Solar Dynamic Observatory (\textit{SDO}) therefore presents a unique opportunity to better understand the AR energetics with its full disk, high resolution and high cadence vector magnetograms.

Here we report the evolution of magnetic field and energy in NOAA AR 11158 using a series of HMI vector magnetograms and a non-linear force free field (NLFFF) extrapolation. This 5-day uninterrupted, 12-minute cadence data set allows us to study in detail both the \textit{long-term}, \textit{gradual} evolution, as well as the \textit{rapid} changes during an X-class flare and CME eruption. We briefly describe the data set and the extrapolation method in Section~\ref{sec:method}. The long-term evolution and the rapid changes are discussed in Sections~\ref{sec:evolution} and~\ref{sec:collapse}, respectively. In addition to the HMI data, we employ coronal and chromosphere images, such as those taken by the Atmospheric Imaging Assembly \citep[AIA;][]{lemen2011} on \textit{SDO} for context and validation of the results.


\begin{figure*}
\centerline{\includegraphics{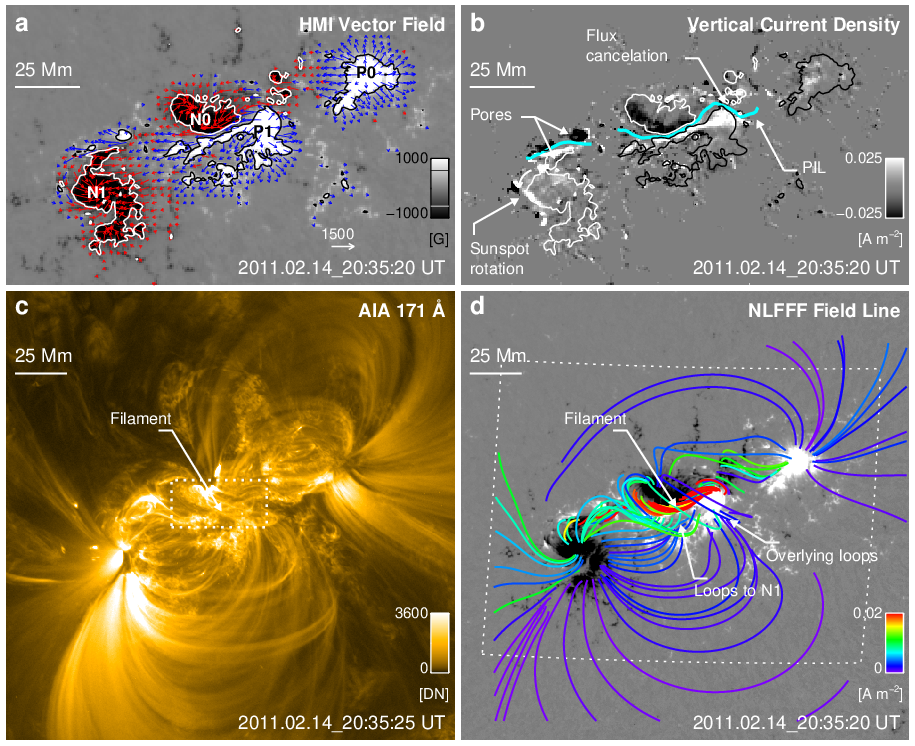}}
\caption{Observations and modeling results for 2011 February 14 20:35 UT, about 5 hours before the X-class flare. (a) Remapped HMI vector magnetogram for the center region of AR 11158 as viewed from overhead. Vertical field ($B_z$) is plotted in the background; blue (red) arrows indicate horizontal field ($B_h$) with positive (negative) vertical component. Contours are plotted at $\pm600\;\rm{G}$. (b) Vertical current density ($J_z$) derived from 5-pixel Gaussian-smoothed vector magnetogram. Contours are for $B_z$ and are identical to (a). Part of the polarity inversion line (PIL) is plotted as thick cyan curve. (c) Image from AIA 171 {\AA} band showing the corona magnetic structures. The dotted box in the center indicates the FOV of Figure~\ref{f:filament}a. (d) Selected field lines from the NLFFF extrapolation plotted over a cutout from the vertical field map. The lines are color-coded by the vertical current density at their footpoints (see the color bar); red field lines correspond to strong current density. The white dotted box indicate the FOV of (a) and (b). The FOVs of (c) and (d) are identical, about $218\times218\;\rm{Mm^2}$, or $302\arcsec\times302\arcsec$. Features of interest are marked in each panel; see text for details. (An animation of the vector field data set for the entire 5-day period is available in the online journal.) \label{f:field}}
\end{figure*}

\section{Vector Magnetograms and Field Extrapolation}
\label{sec:method}

The HMI instrument on \textit{SDO} observes the full Sun at 6 different wavelengths and 6 polarization states in the Fe I 6173 {\AA} absorption line. Filtergrams with 0.5{\arcsec} pixels are collected and converted to observable quantities on a rapid time cadence. For the vector magnetic data stream, each set of filtergrams takes 135 seconds to complete.

To obtain vector magnetograms, Stokes parameters are first derived from filtergrams observed over a 12-minute interval and then inverted through a Milne-Eddington based algorithm, the Very Fast Inversion of the Stokes Vector \citep[VFISV;][]{borrero2011}. The 180$^\circ$ azimuthal ambiguity in the transverse field is resolved by an improved version of the ``minimum energy'' algorithm \citep{metcalf1994,metcalf2006,leka2009}. Regions of interest (ROIs) with strong magnetic field are automatically identified near real time \citep{turmon2010}. A detailed description on the production and relevant characteristics, specifically for the data set for AR 11158, can be found in \cite{hoeksema2012}. Outstanding questions regarding data reduction are also described there.

The magnetic field above the lower chromosphere largely satisfies the force-free criteria \citep{metcalf1995}, where currents align with the magnetic field. For the most general case,
\begin{equation}
\label{eq:nlfff}
\nabla\times\bm{B}=\alpha\bm{B},
\end{equation}
\begin{equation}
\label{eq:divb}
\bm{B}\cdot\nabla\alpha=0,
\end{equation}
with the torsion parameter, $\alpha$, varying in space but constant along each field line. We use an optimization-based NLFFF algorithm \citep{wiegelmann2004} to extrapolate the coronal field from the magnetograms in a Cartesian domain. A preprocessing procedure removes most of the net force and torque from the data so the boundary can be more consistent with the force-free assumption \citep{wiegelmann2006}. For reference, we also construct a potential field (PF) from the same observation using the vertical component of the field and a Green function algorithm \citep[e.g.][]{sakurai1989}. The magnetic free energy ($E_f$) can be inferred by subtracting the PF energy ($E_P$) from the NLFFF energy ($E_N$), where the energy is computed from the field strength within a certain volume $V$:
\begin{equation}
\label{eq:E}
E_f = \int_V \, \frac{B_N^2}{8\pi} \: {\rm d}V - \int_V \, \frac{B_P^2}{8\pi} \: {\rm d}V.
\end{equation}
Here the subscripts $N$ and $P$ denote NLFFF and PF respectively. We note that the VFISV inversion scheme for HMI has the magnetic filling factor held at unity, so the obtained ``magnetic field'' is essentially averaged flux density. In the context of this work, we do not distinguish these two and use the unit Gauss (G) throughout.

We use 600 vector magnetograms of NOAA AR 11158 with a cadence of 12 min, from 2011 February 12 to 16 (see online animation of Fig.~\ref{f:field} for the data set\footnotetext{\url{http://sun.stanford.edu/~xudong/Article/field.mp4}}). The images are de-rotated to the disk center and remapped using Lambert equal area projection \citep{calabretta2002,thompson2006}. The field vectors are transformed to Heliographic coordinates with projection effect removed \citep{gary1990}. For direct analysis of the photospheric field, we use full resolution data at about $360\;\rm{km\,pix^{-1}}$ (0.5{\arcsec}). For extrapolation, we bin the data to $720\;\rm{km\,pix^{-1}}$ (about 1{\arcsec}) and adopt a computation domain of $216\times216\times184\;\rm{Mm}^3$ ($300\times300\times256$).

Because the preprocessed bottom boundary emulates the magnetic field in the chromosphere, we assign a uniform altitude of $720\;\rm{km}$ (1 pixel) to this layer where the field has become largely force-free \citep{metcalf1995} and use it through out the study. In general, we limit the field-of-view (FOV) for analysis to the center $184\times144\times115\;\rm{Mm^3}$, covering most of the strong field region. Uncertainties are reported as mean $\pm$ standard deviation unless specified otherwise.

The extrapolation is performed at 1-h cadence. For the 9 h around the X-class flare the full 12-min cadence is utilized. For convenience, we choose February 15 00:00 UT as time 0 ($T = 0\;\rm{h}$) and use this convention when needed. The AR passed central Meridian on early February 14 ($-22.2\;\rm{h}$). The X2.2 flare (W21S21) started at 1.7 h (February 15 01:44 UT) and peaked at 1.9 h (01:56 UT) in the GOES soft X-ray (SXR) flux and was accompanied by a front-side halo CME \citep{schrijver2011x}.


\section{Long-Term Evolution}
\label{sec:evolution}

\begin{figure*}
\centerline{\includegraphics{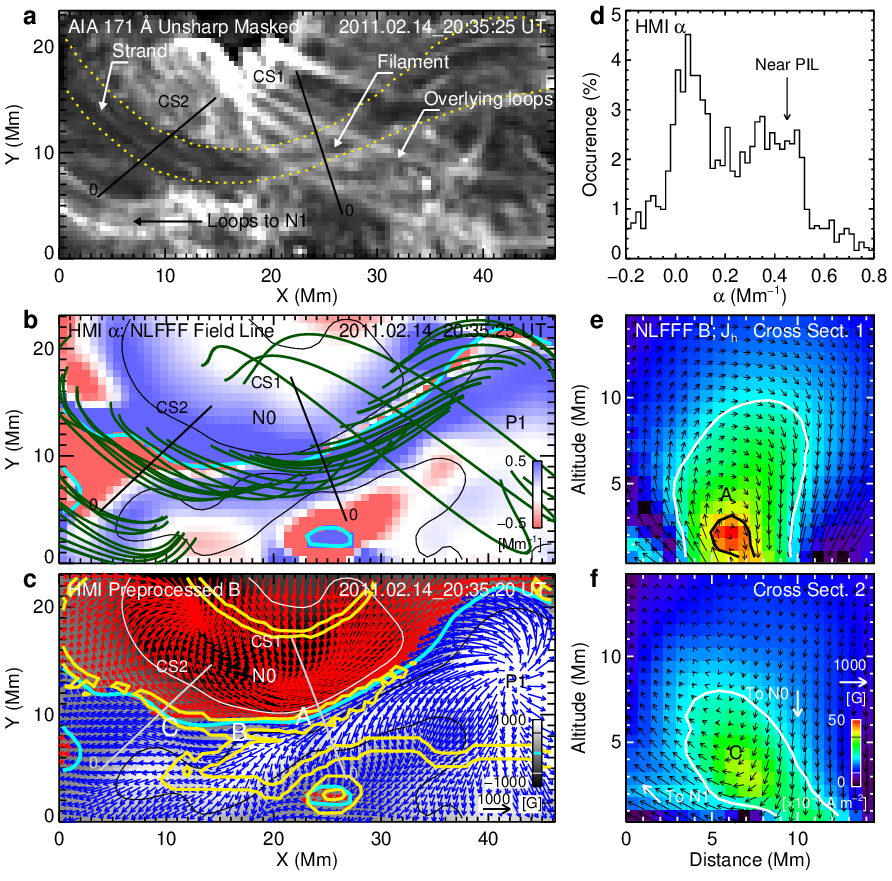}}
\caption{Close-up view of the AR core field structure for the 2011 February 14 20:35:20 UT frame. (a) Unsharp masked AIA 171 {\AA} image. The FOV is denoted by the dotted box in Figure~\ref{f:field}c. Two straight lines show the baselines of the vertical cross sections (CS1 and CS2) through the computation domain plotted in panels (e) and (f). Yellow dotted lines are manually drawn to outline the dark \textit{S}-shaped filament plasma. One of the faint strands appearing in the filament as well as a few bright overlying loops are marked on the image. (b) Selected NLFFF field lines. The torsional parameter, $\alpha$, derived from preprocessed magnetogram is shown as background. Contours are for $B_z=\pm600\;\rm{G}$; the cyan line is for the PIL. (c) Preprocessed, remapped HMI vector magnetogram showing the emulated chromospheric field. The yellow contours indicate the 90$\%$ level of the magnetic connectivity gradient metric, $\tilde{N}$ (equation~\ref{eq:norm} in Appendix~\ref{a:definition}). The contour is an indicator of the intersection of quasi-separatrix layer (QSL) with the lower boundary. A, B, and C mark features of interest. (d) Histogram of $\alpha$ distribution in the FOV of panel (b). The peak near $0.46\;\rm{Mm^{-1}}$ comes from the region along the PIL. (e) Horizontal current density ($J_h$) distribution on a vertical cross section (CS1) with projected NLFFF field vectors. Only the component perpendicular to the cross section is shown for $J_h$. The white and black contours are for 10 and $36\;\rm{mA\,m^{-2}}$ $J_h$, respectively. (f) Similar to (e), showing inverse-polarity configuration (CS2), where field vectors turn upward from right to left. \label{f:filament}}
\end{figure*}

\begin{deluxetable}{clcc}
\tablewidth{0pt}
\tablecaption{Estimated field parameters in the filament channel.\label{t:filament}}
\tablehead{\colhead{Parameter\tablenotemark{a}} & \colhead{Unit} & \colhead{Cross section\tablenotemark{b}} & \colhead{Near axis\tablenotemark{c}}}
\startdata
$A$ & $\rm{Mm^2}$ & 59.9 & 4.2  \\
$\Phi_{a}$ & $10^{20}\;\rm{Mx}$ & 4.87 & 0.49 \\
$\langle B \rangle$ & $10^3\;\rm{G}$ & $0.97\pm0.27$ & $1.20\pm0.05$ \\
$\langle |\gamma| \rangle$ & degree & $26\pm16$ & $11\pm8$ \\
$\langle J_h \rangle$ & $\rm{mA\,m^{-2}}$ & $19\pm9$ & $43\pm5$ \\
$\langle \alpha \rangle$ & $\rm{Mm^{-1}}$ & $0.27\pm0.15$ & $0.46\pm0.05$ \\
$\langle L \rangle$ & $\rm{Mm}$ & $29\pm5$ & $24\pm3$ \\
$\langle \phi \rangle$ & Turns & $0.64\pm0.17$ & $0.92\pm0.17$ 
\enddata
\tablenotetext{a}{All parameters are estimated from uniformly sampled points on a vertical cross section through the NLFFF extrapolation domain, marked as ``CS1'' in Figure~\ref{f:filament}. Angle brackets refers to the mean value. The notations are as follows: cross section area $A$, axial flux $\Phi_a$, field strength $B$, inclination angle $\gamma$, horizontal current density $J_h$, torsional parameter $\alpha$, loop length $L$, twist angle $\phi$. The inclination angle is measured with regard to the photosphere, ranging from $-90^\circ$ to $90^\circ$ with $0^\circ$ for horizontal field and sign consistent with polarity of $B_z$. $\alpha$ is estimated using equation~\ref{eq:alpha}. $\phi$ is divided by $2\pi$ to show the estimated number of turns.}
\tablenotetext{b}{Values are calculated within the $J_h>10\;\rm{mA\,m^{-2}}$ contour in Figure~\ref{f:filament}e and reported as mean $\pm$ standard deviation when applicable.}
\tablenotetext{c}{Values are calculated within the $J_h>36\;\rm{mA\,m^{-2}}$ contour; the centroid is near $3\;\rm{Mm}$ in height.}
\end{deluxetable}

\subsection{An Illustrative Snapshot}
\label{subsec:snapshot}

We use a snapshot taken at 20:35 UT on February 14, about 5 h before the X-class flare, to illustrate the magnetic field structures in AR 11158 when it is well developed. It mainly consists of two bipoles (P0/N0 and P1/N1 in Fig.~\ref{f:field}a, P denoting positive and N negative) that form a complex quadrupolar structure \citep{schrijver2011x}. The total unsigned magnetic flux ($|\Phi|$) is about $2.7\times10^{22}\;\rm{Mx}$ and the flux is balanced to within $1\%$. Vertical field ($B_z$) in the center of umbra can be as strong as $2600\;\rm{G}$. Strong shearing motion exists between P1/N0 as well as amongst a few newly emerged pores. The horizontal field ($B_h$) in the AR center is largely parallel to the major polarity inversion line (PIL; Fig.~\ref{f:field}a), near which strong vertical electric currents ($J_z$) are present (Fig.~\ref{f:field}b). On the south side of the PIL, an elongated strip of positive flux following P1 forms a so-called ``magnetic tongue'' \citep[e.g.][]{luoni2011}. Fast sunspot rotation and flux cancellation are also observed.

AIA 171 {\AA} observation of the extreme-ultraviolet (EUV) light emitting plasma roughly outlines the coronal magnetic field (Fig.~\ref{f:field}c). As a validation of our field extrapolation, we plot selected field lines for comparison and color-code them according to $|J_z|$ at their footpoints (Fig.~\ref{f:field}d). Brighter color, such as red, corresponds to stronger $|J_z|$. The modeled field lines with strong currents indicate non-potential structures, and are qualitatively in good agreement with observation. In particular, field lines rooted in the core region (regions near the major PIL) morphologically resemble the observed EUV filament. In contrast, potential-like loops (with weaker current) further away from the center are not well recovered, especially on the north side. A more detailed discussion on the extrapolation can be found in Appendix~\ref{a:extrapolation}.

A closer look at the core region, summarized in Figure~\ref{f:filament}, reveals a close match between the shape of the filament and the photospheric PIL. There seem to be a few faint strands in the dark filament: NLFFF extrapolation suggests that an ensemble of highly sheared loops thread the plasma. These loops are rooted in P1/N0 inside a narrow strip of strong-current, high-$\alpha$ distribution along the PIL. These loops are typically low-lying, with apexes well below $10\;\rm{Mm}$. We note the broad distribution of photospheric $\alpha$, mainly from 0 to $0.5\;\rm{Mm^{-1}}$ (Fig.~\ref{f:filament}d), necessitating the non-linear treatment of any realistic modeling attempts.

Figure~\ref{f:filament}e shows in the background the computed horizontal current density ($J_h$) in a vertical plane perpendicular to the PIL (near ``A'' in panel c). The projected magnetic field vectors on this cross section rotate around an axis at about $3\;\rm{Mm}$ altitude; $J_h$ peaks at the same height. Signatures of twisted flux ropes, such as an X-point or a hollow core $J_h$ distribution around its axis \citep[e.g.][]{bobra2008,su2011} are not obvious.

\begin{figure*}
\centerline{\includegraphics{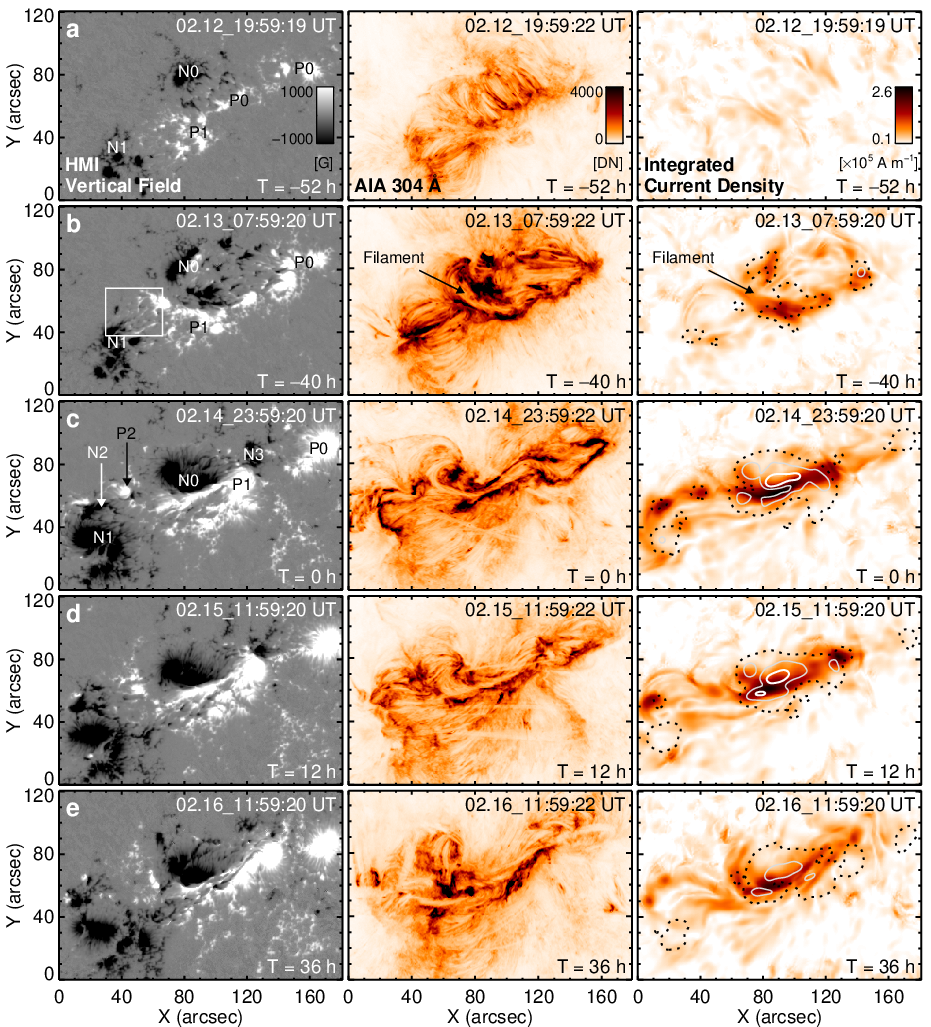}}
\caption{Five snapshots of the evolving AR11158. They are taken at about $T=-52\;\rm{h}$, $-40\;\rm{h}$, $0\;\rm{h}$, $12\;\rm{h}$, and $36\;\rm{h}$, with February 15 00:00 UT as time 0. Left column: HMI $B_z$ as in native coordinate (as recorded by camera). Middle column: negative AIA 304 {\AA} image showing chromosphere and transition region structures in which the AR filament is best discernible. Right column: vertically integrated current density from NLFFF extrapolation over the lowest $10\;\rm{Mm}$. The thick solid, thin solid, and dotted contours are for similarly integrated free energy density at $80\%$, $50\%$, and $20\%$ of the peak value for frame $T=0\;\rm{h}$. Images are remapped back to the native coordinate for direct comparison of HMI and AIA observations. The box in (b) indicates the field-of-view of Figure~\ref{f:emergence} for a flux-emerging region. Features of interest are marked in some panels; see text for details. (An animation of the entire 5-day period is available in the online journal.) \label{f:evo}}
\end{figure*}

Using AIA 304 {\AA} images (Fig.~\ref{f:evo}) where the filament plasma is best defined, we estimate the width of the EUV filament to be about $6\;\rm{Mm}$, slightly wider than it appears in the 171 {\AA} band. The apparent length is at least $90\;\rm{Mm}$. We choose the $10\;\rm{mA\,m^{-2}}$ $J_h$ contour (Fig.~\ref{f:filament}e) as a proxy for the filament cross section as it encloses a region comparable to the observed filament width. A summary of the estimated field parameters in this region (filament channel) are listed in Table~\ref{t:filament}. In particular, the field strength ($B$) near the $J_h$ maximum is about $1200\;\rm{G}$. This inferred kilogauss field is strong compared with previously modeled plage filaments \citep[e.g.][]{aulanier2003,guo2008,jing2010} and recent observation \citep[e.g.][]{kuckein2009}. We estimate the field twist angle as $\phi=\alpha L/2$ \citep{longcope1998} with $L$ being the loop length and find an average of 0.6 turn, with about 0.9 turn near the $J_h$ maximum.

The field line mapping appears to ``bifurcate'' near the PIL on the east side (``B'' in Fig.~\ref{f:filament}c). Some EUV loops originate from the P1 magnetic tongue connect back to N0; others deviate from the PIL and connect to N1 more than $50\;\rm{Mm}$ away (Fig.~\ref{f:field}c and \ref{f:filament}a). The photospheric $\alpha$ changes sign here as well. We exploit the extrapolation and calculate the magnetic connectivity gradient metric $\tilde{N}$ (equation~\ref{eq:norm} in Appendix~\ref{a:definition}) on the lower boundary \citep{demoulin1996}. High-$\tilde{N}$ contours show the photospheric intersection of the quasi-separatrix layers (QSLs). They divide the magnetic tongue into two distinctive regions (Fig.~\ref{f:filament}c), where the computed field connectivities diverge.

In addition, an interesting ``inverse-polarity'' configuration \citep[e.g.][and references therein]{mackay2010} exists nearby too, where horizontal field on the photosphere points from negative vertical polarity side to the positive (near ``C'' in Fig.~\ref{f:filament}c). On the vertical cross section shown in Figure~\ref{f:filament}f, projected field vectors form a ``concave-up'' configuration. Loops turn upward here or become nearly parallel to the photosphere.

\subsection{Evolution of Magnetic Field}
\label{subsec:field}

We show five representative snapshots of the evolving AR in Figure~\ref{f:evo} (see online animation for the entire 5-day evolution with full cadence\footnotetext{\url{http://sun.stanford.edu/~xudong/Article/evo.mp4}}). The left and middle columns show $B_z$ and negative AIA 304 {\AA} images, respectively. The right column shows the vertical integration of absolute current density ($J$) over the lowest $10\;\rm{Mm}$ as derived from the NLFFF extrapolation. Patterns of strong current concentration may serve as a proxy for the non-potential coronal structures \citep[e.g.][]{schrijver2008nlfff}.

\begin{figure*}
\centerline{\includegraphics{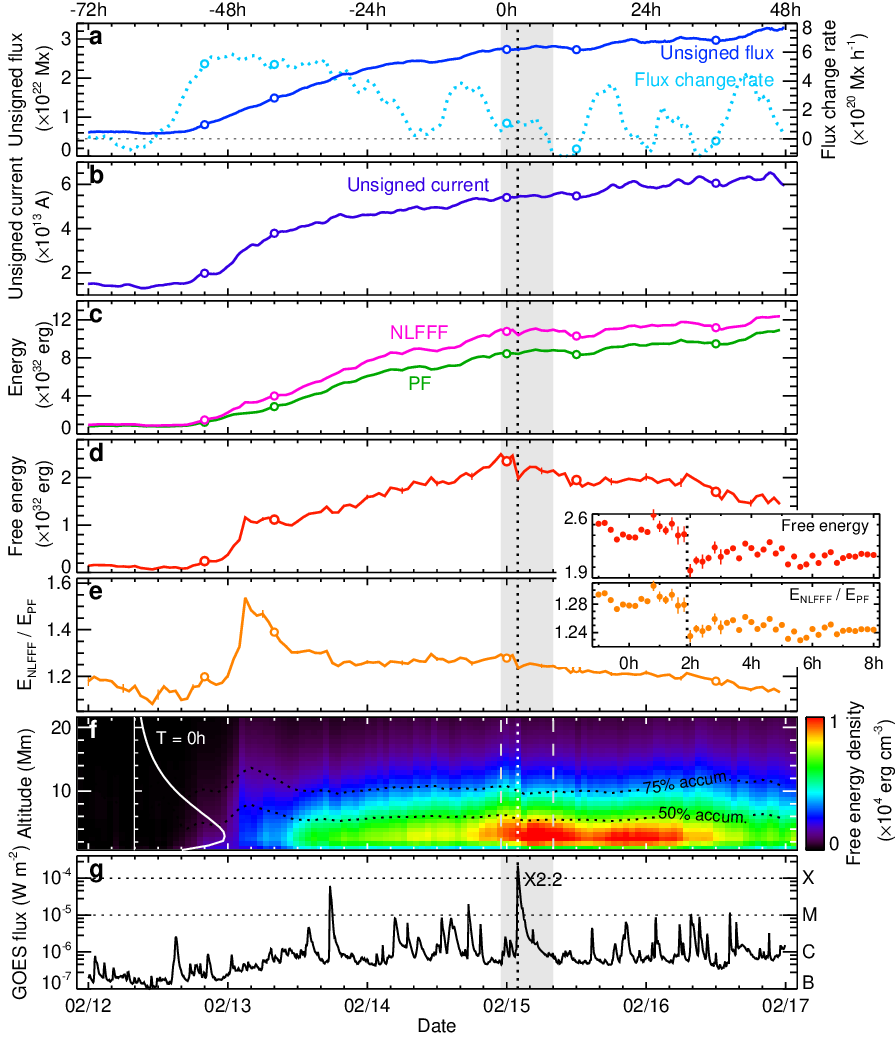}}
\caption{Evolution of magnetic energy and related quantities of AR 11158 over 5 days. (a) Total unsigned magnetic flux ($|\Phi|$) and 6-h smoothed flux change rate (${{\rm d}|\Phi|}/{{\rm d}t}$). (b) Total unsigned current ($|I|$). (c) Magnetic energy derived from the NLFFF and PF extrapolation ($E_N$ and $E_P$). (d) Estimated magnetic free energy ($E_f$). (e) Ratio between the NLFFF and PF energy ($E_N/E_P$). (f) Time-altitude diagram of average magnetic free energy density in the AR center (FOV of Fig.~\ref{f:flare}(f)). Dotted lines indicate the height below which the accumulated values reach $50\%$ and $75\%$ of the total. Curve on the left shows the height profile for $T=0\;\rm{h}$. (g) GOES 5 m soft-X ray flux (1--8 {\AA} channel). The 5 frames in Figure~\ref{f:evo} are marked as circles.  Flux and current are derived using only pixels with $|B_z|$ greater than $100\;\rm{G}$. Error bars in (c)-(e) show the estimated effect of noise and are plotted every 6 h, but are usually too small to be seen. Insets of (d) and (e) show results with a 12 minute cadence from -1 h to 8 h (shaded grey band). Error bars right before and after the X-class flare in these insets are more visible. The vertical dotted lines indicate the peak time of the X-class flare at 1.9 h.\label{f:energy}}
\end{figure*}

Here we highlight a few key features of the magnetic field evolution. For reference, the total unsigned flux $|\Phi|$ and its change rate (${{\rm d}|\Phi|}/{{\rm d}t}$) are plotted in Figure~\ref{f:energy}a; the photospheric unsigned current ($|I|$) in Figure~\ref{f:energy}b; the GOES SXR flux in Figure~\ref{f:energy}g.

\begin{itemize}
\renewcommand{\labelitemi}{$-$}
\item The early stage of the AR development (top two rows of Fig.~\ref{f:evo}) features fast flux emergence that coincides with the appearance of a pronounced filament. Such fast emergence continues for about 1 day (Fig.~\ref{f:energy}a) with a rate of several $10^{20}\;\rm{Mx\,h^{-1}}$, mainly in the aforementioned two bipoles P0/N0 and P1/N1 \citep{schrijver2011x}. Note that N0 consists of two sunspots, the western one emerges much later. AIA images show frequent brightening as P1 and N0 converge and collide; new magnetic connectivities are being established rapidly in the corona. A filament becomes visible along the PIL within a few hours. Along the PIL, significant electric current injection appears to take place (Figs.~\ref{f:evo}b and \ref{f:energy}b). During this interval, the net (unbalanced) current in each magnetic polarity also increases drastically, hinting that the newly emerged flux is highly non-potential \citep{schrijver2009}.

\item After the initial fast flux emergence (third row of Fig.~\ref{f:evo}), the sunspot complex develops further with slower flux emergence but lasting strong shearing motion between P1 and N0. The injection of current through the photosphere clearly slows down (Fig.~\ref{f:energy}b). However, the vertically integrated $|J|$ still increases by about $25\%$ over a period of 40 h, suggesting a net build-up process in the corona. A pair of newly emerged pores (P2/N2 in Fig.~\ref{f:evo}c) with a few $10^{20}\;\rm{Mx}$ flux undergo fast shearing on the northeastern side and substantially reconfigure the coronal field. Smaller recurrent eruptions take place above these pores throughout the next two days. At the same time, the filament appears to be stretched. It increases in length, becomes somewhat warped (perhaps elevated) on the western end and extends further west to P0 (Fig.~\ref{f:evo}c). Simultaneous flux cancellation is observed between P1 and N3, which may act favorably to the subsequent eruptions.

\item Toward the end of the 5-day period (bottom two rows of Fig.~\ref{f:evo}), flux emergence becomes episodic, sometimes overtaken by flux cancellation (Fig.~\ref{f:energy}a). Nevertheless, the shearing between N0 and P1 continues. The integrated $|J|$ in the AR core region shows little sign of decreasing, and is fluctuating within $10\%$ of the pre-flare value. A similar trend is present in $|I|$ as well (Fig.~\ref{f:energy}b). Over time, the converging motion between sunspots of the same polarity starts to simplify the region toward a more bipolar structure. The filament seems to survive multiple eruptions and is still visible at the end of the 5 days.
\end{itemize}

\subsection{Evolution of Magnetic Energy}
\label{subsec:efree}

We now consider the distribution of the magnetic free energy based on the NLFFF extrapolation. Spatially, the low-lying, current-carrying core field demonstrates strong concentration of free energy in the AR core, from the chromosphere to the lower corona. The free energy density ($\epsilon_f$) appears largely co-spatial with the current distribution (Fig.~\ref{f:evo} right column). At $T=0\;\rm{h}$, the $20\%$ level contour of the peak $\epsilon_f$ (vertically integrated) accounts for only $9\%$ of the AR area (Fig.~\ref{f:evo}c), but $77\%$ of the overall free energy. The height profile of $\epsilon_f$ plotted in Figure~\ref{f:energy}f shows that about $50\%$ of all free energy is stored below $6\;\rm{Mm}$, and $75\%$ below $11\;\rm{Mm}$. We note that the maximum free energy density occurss at about $3\;\rm{Mm}$ altitude, corresponding to the height of the peak in $J_h$ (Fig.~\ref{f:filament}e). These distribution patterns remain relatively stable after the filament becomes well defined in AIA images around -36 h.

The temporal profiles of the total NLFFF and PF magnetic energy ($E_N$ and $E_P$) in Figure~\ref{f:energy}c roughly scale with the unsigned flux. On the other hand, the estimated magnetic free energy $E_f$ in Figure~\ref{f:energy}d shows interesting variations in time and appears sensitive to AR activity, as one would expect. We plot the ratio $E_N/E_P$ in Figure~\ref{f:energy}e as an additional measurement of the AR's non-potentiality. Uncertainties reported in this section only represent the effect of spectropolarimetric noise and are obtained from a pseudo Monte-Carlo method (see Section~\ref{subsec:energy}). Systematic uncertainties from the extrapolation algorithm may be greater. Again, several key features are highlighted below.

\begin{itemize}
\renewcommand{\labelitemi}{$-$}
\item \textit{Rapid free-energy injection during flux emergence.} $E_f$ increases drastically following the flux emergence (Fig.~\ref{f:energy}a and b). Free energy concentration is seen in the vicinity of the filament (Fig.~\ref{f:evo}b), and $\epsilon_f$ at all altitudes increases significantly (Fig.~\ref{f:energy}f). In particular, $E_f$ shows a twelve-fold increase from -55 h to -45 h, reaching $1.07\times10^{32}\;\rm{erg}$ and amounting to $43\%$ of the final maximum $E_f$ two days later. The ratio of $E_N/E_P$ peaks at about 1.57 at -45 h. Significant changes in the coronal structures ensue, as conveyed by the AIA images (animation for Fig.~\ref{f:evo}). Flux change rate and $E_f$ then plateau for a few hours, and $E_N/E_P$ drops back to about 1.30. Careful inspections of the vector magnetogram reveals signatures of emerging flux tubes, which we discuss in Section~\ref{subsec:energy}.

\item \textit{Gradual energy build-up and decay.} After the initial fast increase, $E_f$ accumulates at a slower rate (Fig.~\ref{f:energy}d). The AR has attained a well-developed quadrupolar topology. The growth rate of $E_f$ and the ratio $E_N/E_P$ are both nearly constant. Similar trends can be found for the post X-flare stage, except now $E_f$ starts to decrease and the region relaxes to a more potential state (Fig.~\ref{f:energy}d and e). The energy dissipation rate in the corona probably has exceeded the growth rate from the photosphere as the flux emergence slows down (Fig.~\ref{f:energy}a).

\item \textit{Step-wise energy loss during the X-class flare.} $E_f$ reaches $(2.47\pm0.03)\times10^{32}\;\rm{erg}$ prior to the X-class flare, accounting for about $23\%$ of the total energy. During the flare, $E_f$ displays a step-wise sudden decrease (Fig.~\ref{f:energy}d and inset). The 1 h average of $E_f$ before and after the flare differ by about $(0.34\pm0.04)\times10^{32}\;\rm{erg}$, $14\%$ of the pre-flare $E_f$. This value situates at the lower end of what is adequate to power a typical X-class flare \citep[e.g.][and references therein]{hudson2011} and may be intrinsically an underestimation, as discussed in Section~\ref{subsec:energy}. A clear discontinuity also appears in the $E_N/E_P$ and $\epsilon_f$ height profile. Sudden energy decrease is also found during the earlier M-6.6 flare (-30.4 h) with a smaller magnitude. 

\end{itemize}

The continuous monitoring of the AR free energy with relatively high temporal/spatial resolution is made possible, for the first time, by the HMI vector field observations. This is especially useful for the study of major eruptive events. Due to the nature of the extrapolation method, changes in $E_f$ here are determined by the boundary conditions. Therefore, the step-wise energy loss found during the X-class flare must be related to rapid and significant field changes on the photosphere. We consider this change in detail in Section~\ref{sec:collapse}.


\section{Flare-Related Rapid Change}
\label{sec:collapse}

\begin{figure*}
\centerline{\includegraphics{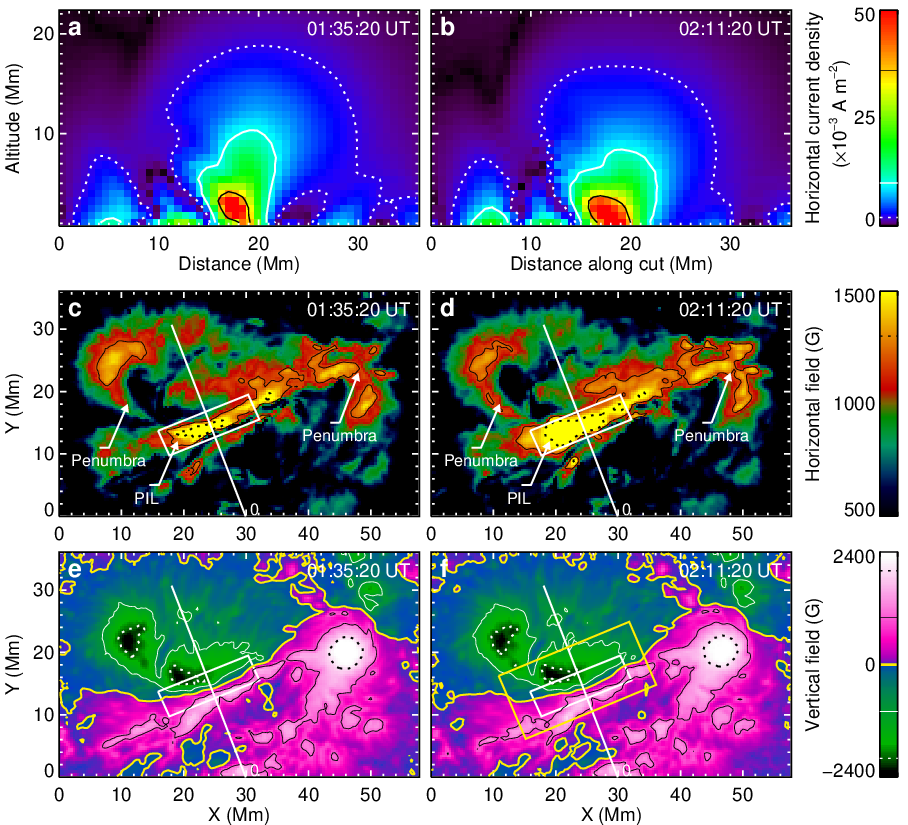}}
\caption{Rapid field changes at the photosphere and in the corona during the 2011 February 15 X-class flare. (a) $J_h$ distribution on a vertical cross section as derived from NLFFF extrapolation, before the flare at 01:35:20 UT. Only the component perpendicular to the cross section is included. The location of the cross section is indicated in panels (c)-(f) as a long straight line. The dotted, dashed and solid contours indicate values of 2, 10, and 36 $\rm{mA\,m^{-2}}$ respectively. (b) Same as (a), for 02:11:20 UT after the flare. (c) Remapped $B_h$ observed by HMI for 01:35:20 UT as viewed from overhead. The dotted (solid) lines show contour for $1600\;\rm{G}$ ($1200\;\rm{G}$). Places with significant field change are marked by arrows and boxes. (d) Same as (c), for 02:11:20 UT. (e) Remapped $B_z$ at 01:35:20 UT. The dotted (solid) contours are for $\pm2000\;\rm{G}$ ($\pm1000\;\rm{G}$). (f) Same as (e), for 02:11:20 UT. The large yellow box indicates the region evaluated in Figure~\ref{f:relax}(d). (An animation of this figure is available in the online journal.) \label{f:flare}}
\end{figure*}

\begin{figure*}
\centerline{\includegraphics{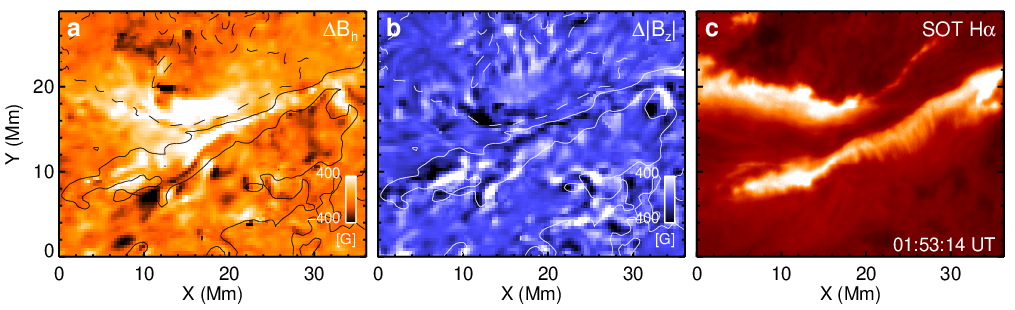}}
\caption{Difference images of pre- and post-flare magnetic field. (a) Difference of $B_h$ between 02:11:20 UT and 01:35:20 UT. The two frames are kept in the native coordinates without remapping, but are co-aligned with Carrington rotation rate to the nearest full pixel. Contours are for the mean $B_z$ of the two frames, at $\pm600\;\rm{G}$. (b) Difference image of $|B_z|$. (c) H$\alpha$ flare ribbons observed by \textit{Hinode}/SOT at 01:53:14 UT, near flare peak and roughly midway between the two HMI frames. \label{f:ribbon}}
\end{figure*}
\begin{figure*}
\centerline{\includegraphics{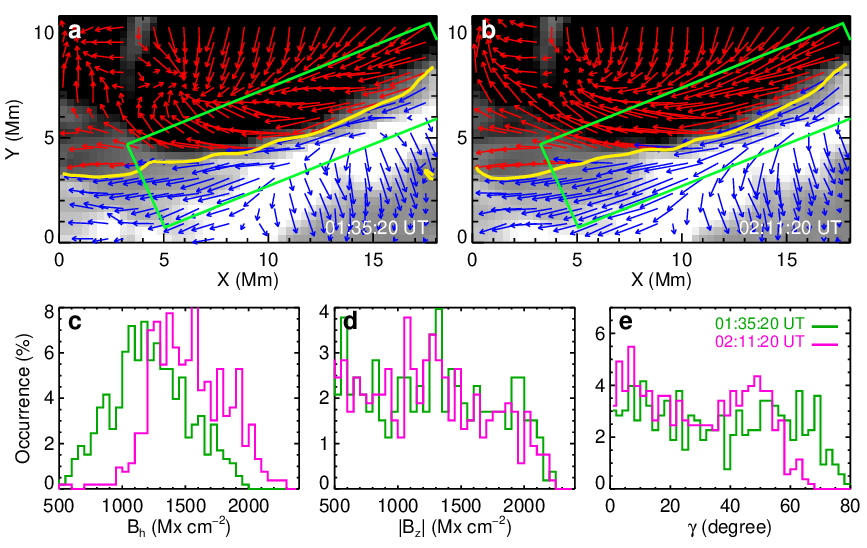}}
\caption{Close-up view of the pre- and post-flare vector field in the AR core region. (a) Remapped vector magnetogram for 01:35:20 UT. The yellow contour is the PIL; the boxed region is identical to Fig.~\ref{f:flare}c. (b) Same as (a), for 02:11:20 UT. (c) Histograms of $B_h$ at two different times, in the boxed region in panel (a). (d) Histograms for $|B_z|$. (e) Histograms for inclination angle $|\gamma|$. \label{f:dvec}}
\end{figure*}
\begin{figure*}
\centerline{\includegraphics{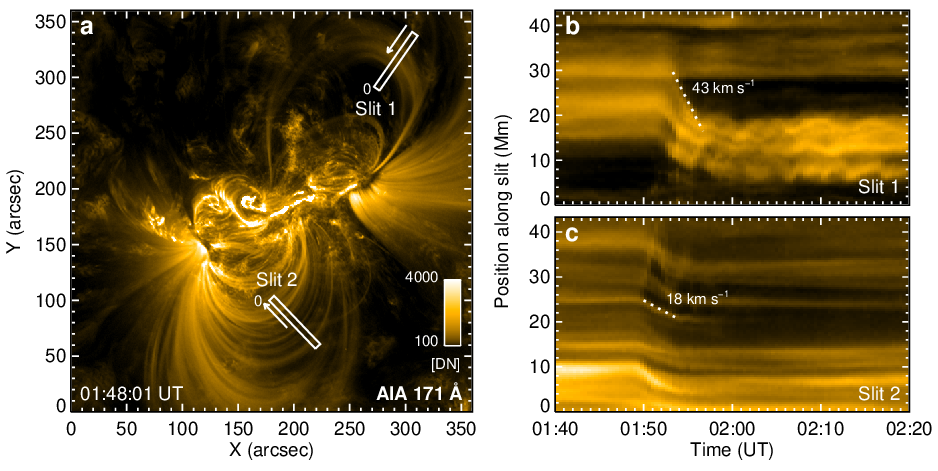}}
\caption{Observed EUV coronal loop retractions. (a) AIA 171 {\AA} image at 01:48:01 UT on February 15, after the onset of flare. Two slits that are largely perpendicular to the loops are used to obtain the time-position diagrams in the following panels. Arrows along the slits indicate the approximate direction of the transverse motion. (b) Time-position diagram for slit 1 constructed by stacking a time sequence of coaligned image slices from left to right. Only the images with the normal exposure time are used, resulting in a 24 s cadence. Images are coaligned to sub-pixel accuracy. The dotted line shows an inward loop contraction pattern with a transverse speed of $43\;\rm{km\,s^{-1}}$. The horizontal patterns in the upper half are from the features in the background. Loop oscillations are visible. (c) Same as (b), for slit 2. Dotted line shows a transverse speed of $18\;\rm{km\,s^{-1}}$. Patterns of expanding loops also appear around 01:50 UT, moving from position 20 Mm toward 0. (An animation of this figure is available in the online journal.) \label{f:contract}}
\end{figure*}

\begin{figure*}
\centerline{\includegraphics{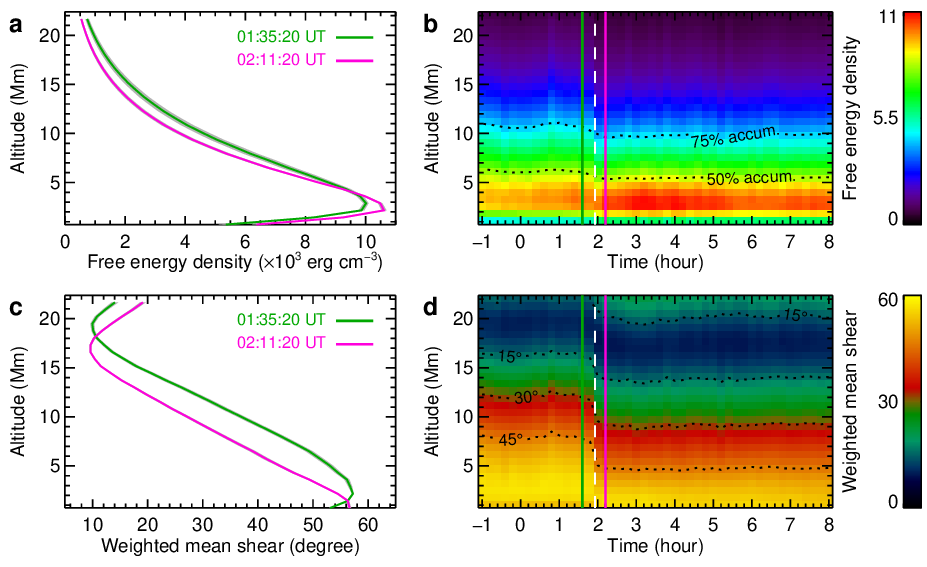}}
\caption{Altitude profile for extrapolated field around the time of the X-class flare. (a) Profiles of average magnetic free energy density in the AR core region (FOV of Figure~\ref{f:flare}f) at February 15 01:35:20 UT and 02:11:20 UT. Shaded grey band indicates the uncertainty due noise, and is usually to narrow to be seen. (b) Time-altitude diagram for free energy density during the 9 h around the X-class flare, similar to Fig.~\ref{f:energy}f. The vertical dashed line indicates the flare peak time at 1.9 h. Green and pink vertical lines indicate the two profiles shown in panel (a). (c) Profiles of weighted mean shear ($\theta_w$) near the PIL (boxed region in Figure~\ref{f:flare}f), at February 15 01:35:20 UT and 02:11:20 UT. (d) Time-altitude diagram for $\theta_w$. Dotted lines show the heights at which the shear reaches $15^\circ$,  $30^\circ$, and $45^\circ$ respectively. PF becomes mostly east-west directed at about $20\;\rm{Mm}$, a height much lower than that of NLFFF, thus resulting in the $\theta_w$ increase above 20 Mm. \label{f:relax}}
\end{figure*}

Figure~\ref{f:flare} illustrates the rapid magnetic field changes during the X-class flare\footnotetext{\url{http://sun.stanford.edu/~xudong/Article/flare.mp4}}. Strong and permanent enhancement of $B_h$ in the AR core was reported by \citet{wangs2011} based on HMI linear polarization signal, and is confirmed here by full Stokes inversion. From 01:35 UT to 02:11 UT, the mean $B_h$ along PIL (Fig.~\ref{f:flare}c and d) increases from about $1200\;\rm{G}$ to over $1500\;\rm{G}$, $28\%$ within 0.6 h. Changes in $B_z$ appear less significant (Fig.~\ref{f:flare}e and f): mean $|B_z|$ decreases by about $5\%$. We skip the two frames in between (01:47 and 01:59 UT) to avoid possible artifacts from flare emission (see Section~\ref{subsec:implosion}).

Shown in Figure~\ref{f:ribbon}, the difference image of pre- and post-flare $B_h$ bears a striking resemblance to the H$\alpha$ flare ribbons observed by the Solar Optical Telescope \citep[SOT;][]{tsuneta2008} on the \textit{Hinode} satellite \citep{kosugi2007}. The elongated regions that are swept by the evolving ribbons or lying in between show significant $B_h$ enhancement, whereas patches with decayed $B_h$ appear on both the north and south sides. In the narrow region along the PIL, distribution of $B_h$ shifts by about $300\;\rm{G}$ in the histogram (Fig.~\ref{f:dvec}c), displaying a strong boost in the kilogauss range. On the other hand, $|B_z|$ often decreases where $B_h$ increases, but the signals are weaker and appear to be mixed with the opposite. The distribution of $|B_z|$ gently decays in the strongest part, but otherwise remains similar (Fig.~\ref{f:dvec}d). 

\begin{deluxetable}{clcccl}[h]
\tablewidth{0pt}
\tablecaption{Flare-related change in magnetic energy and photospheric field.\label{t:flare}}
\tablehead{\colhead{Parameter\tablenotemark{a}} & \colhead{Unit} & \colhead{Pre-flare\tablenotemark{b}} & \colhead{Post-flare\tablenotemark{b}} & \colhead{Difference\tablenotemark{c}}}
\startdata
$E_f$ & $10^{32}\rm{erg}$ & $2.47\pm0.03$ & $2.13\pm0.03$ & $-0.34\pm0.04$ \\
$E_N/E_P$ & - & $1.29\pm0.00$ & $1.25\pm0.00$ & $-0.04\pm0.01$ \\
$\langle B_h \rangle$ & $10^3\;\rm{G}$ & $1.20\pm0.02$ & $1.53\pm0.01$ & +$28\%$ \\
$\langle |B_z| \rangle$ & $10^3\;\rm{G}$ & $0.96\pm0.03$ & $0.91\pm0.03$ & -$5\%$ \\
$\langle B \rangle$ & $10^3\;\rm{G}$ & $1.66\pm0.01$ & $1.87\pm0.01$ & $+13\%$ \\
$\langle |\gamma| \rangle$ & degree & $36\pm1$ & $29\pm1$ & $-19\%$ \\
$\langle |J_z| \rangle $ & $\rm{mA\,m^{-2}}$ & $36\pm2$ & $27\pm1$ & $-25\%$ \\
$\langle \alpha \rangle$ & $\rm{Mm^{-1}}$ & $0.48\pm0.03$ & $0.30\pm0.02$ & $-37\%$
\enddata
\tablenotetext{a}{$E_f$ and the ratio $E_N/E_P$ are computed using the center region ($184\times144\times115\;\rm{Mm^3}$) of the extrapolation domain. Other parameters are for the AR core photospheric field within the boxed region in Figure~\ref{f:flare}c. Notations are same as Table~\ref{t:filament}.}
\tablenotetext{b}{For $E_f$ and $E_N/E_P$, pre-flare values are calculated from the average of 5 frames between 00:47 to 01:35; post-flare between 02:11 and 02:59. Uncertainties are estimated effect from noise. For all others, pre-flare values are calculated for the 01:35 frame; post-flare for 02:11. Values are reported as mean $\pm$ standard error for comparison.}
\tablenotetext{c}{Difference for $E_f$ or $E_N/E_P$ is the absolute value, others are percentage differences.}
\end{deluxetable}

The combined effect is that the field becomes overall stronger and more inclined in the AR core (Fig.~\ref{f:dvec}e). In addition, the azimuth of $B_h$ appear to change in a fashion such that they become better aligned and more parallel to the PIL in its vicinity (Fig.~\ref{f:dvec}a and b), consistent with previous reports \citep[e.g.][]{wanghm1994}. After the flare, $\alpha$ appears to be smaller, suggesting the field is perhaps less twisted. A summary of the pre- and post-flare field parameters is provided in Table~\ref{t:flare}.

These fast changes suggest a scenario in which the change of coronal connectivity, probably induced by reconnection, feeds back to the photosphere. In particular, newly reconnected loops with footpoints located in the flare ribbons, both shorter and more parallel to the PIL, are \textit{a priori} consistent with the observations. We discuss the topic further in Section~\ref{subsec:implosion}. Morphologically, the changes are in line with the conjectured magnetic ``implosion'' \citep{hudson2000}: a decrease in coronal magnetic energy during explosive events should lead the coronal field to contract, resulting in a ``more horizontal'' photospheric field \citep{hudson2008}. Here we analyze a series of NLFFF snapshots and monitor the $J_h$ distribution on a vertical cross section through the computation domain (Fig.~\ref{f:flare}a and b). The patterns show an apparent contracting motion at flaring time owing to the altered boundary condition (see online animation). The apex of the outermost $J_h$ contour lowers by about $3\;\rm{Mm}$ in Figure~\ref{f:flare}b.

Motions of coronal loops farther away from the AR core provide evidence for the conjecture. During flares and CMEs, a depletion of magnetic energy leads to smaller magnetic pressure gradient, and loops must contract to reach a new balance \citep[e.g.][]{liur2010}. We show AIA 171 {\AA} observation for this event in Figure~\ref{f:contract}. Expansion of the coronal structure and faint circular propagating fronts are visible starting around 01:48 UT, probably linked to CME initiation \citep{schrijver2011x}. At about 01:50 UT, the southern potential-like loops suddenly move inward; the northern loops follow immediately (see online animation\footnotetext{\url{http://sun.stanford.edu/~xudong/Article/contract.mp4}} for Figure~\ref{f:contract}). The retraction proceeds for about 5 minutes, mostly during the flare impulsive phase. We place two slits near the loop apexes to construct time-position diagrams (Fig.~\ref{f:contract}b and c). The inward motion shows a projected speed on the order of tens of kilometers per second, some with apparent oscillations. The transverse displacement can be as large as 15 Mm. The possible ambiguities in these observations are discussed and resolved in Section~\ref{subsec:implosion}.

According to NLFFF extrapolation, such contraction yields a more ``compact'' energy distribution and a coronal field that relaxes more rapidly with height. We show in Figure~\ref{f:relax}a two altitude profiles of the mean free energy density before and after the flare. While the total magnetic free energy decreases after the flare, a larger percentage gets stored in the lower altitudes. Higher overlying field thus becomes even less energetic. Similar trends can be found in the field-weighted mean shear $\theta_w$ (equation~\ref{eq:wshear} in Appendix~\ref{a:definition}). This quantity measures the field-weighted direction difference between the NLFFF and PF field, thus providing a quantitative description of the non-potentiality of the magnetic field. As shown in Figure~\ref{f:relax}c, the whole profile moves downwards by about $2\;\rm{Mm}$, equivalent to a shear decrease at almost all heights. The bottom layer, as an exception, becomes slightly more sheared because $B_h$ is now more parallel to the PIL \citep[cf.][]{wanghm1994}. The time-height diagrams covering 9 h around the flare with 12 m cadence more clearly convey the sudden and permanent nature of the changes (Fig.~\ref{f:relax}b and d).


\section{Discussion}
\label{sec:discussion}

\begin{figure*}
\centerline{\includegraphics{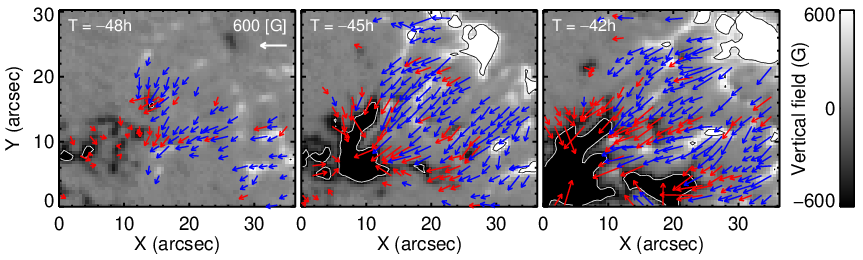}}
\caption{Three snapshots of a flux emergence site. The snapshots are 3 h apart. The FOV corresponds to the boxed region in Figure~\ref{f:evo}b. $B_z$ is plotted in the background; contours are for $\pm600\;\rm{G}$. Arrows indicate highly inclined $B_h$. For clarity, arrows are drawn only on every third pixel, for field strength $200\;\rm{G}<B<800\;\rm{G}$, and for inclination $|\gamma|<20^\circ$. Images are in native coordinates. \label{f:emergence}}
\end{figure*}

\subsection{On the Coronal Energy Budget}
\label{subsec:energy}

As discussed in Section~\ref{sec:evolution}, early flux emergence brings along a large amount of electric current and free energy. We show one of the flux emergence sites in Figure~\ref{f:emergence}. Highly inclined fields (inclination angle $\gamma$ measured from the photosphere less than $20^\circ$) appear within a widening patch, separating two vertical flux concentrations of opposite polarity that themselves grow in size and field strength in time. These patterns appears consistent with an emerging flux-tube at its initial stage of rising into the solar atmosphere.

Before -45 h (February 13 03:00 UT), $E_N$ increases rapidly with the current injection from the increasing $B_h$. These highly inclined fields do not cause vertical flux to increase proportionally. Since the extrapolated potential field solely depends on $B_z$, the growth of $E_P$ lags behind. This may explain the early peak in $E_N/E_P$ shown in Figure~\ref{f:energy}e. The decrease of $E_N/E_P$ can in turn be explained by the subsequent fast increase in vertical flux relative to the horizontal counterpart when the axis of the flux tube approaches the photosphere.

We estimate the effect of spectropolarimetric noise on the energy estimation by conducting a pseudo Monte-Carlo experiment. In general, the noise level for the inverted HMI longitudinal field is below $10\;\rm{G}$; it is about an order higher for the transverse component \citep{hoeksema2012}. For each frame, we create 10 extra-noisy Stokes profiles by adding to them a HMI-characteristic level of photon noise (on the order of $0.1\%$ of quiet Sun intensity). Inversion, disambiguation and modeling are subsequently performed on these 10 inputs, and the standard deviation is adopted. The estimated noise-related uncertainty in $E_N$ is typically around $0.6\%$, accounting for $2\%$--$4\%$ of $E_f$ (Fig.~\ref{f:energy}). This result is consistent with earlier tests on analytical field configurations and synthetic spectra with larger run numbers \citep{tiwari2009}. The relatively small effect is probably owing to the preprocessing procedure, which ensures force-freeness by smoothing away the spurious components. We nevertheless caution that the percentage uncertainty may be above $10\%$ for $E_f$ during the early stage of AR development, when the field is weaker and the measurement signal-to-noise (S/N) is lower.

\begin{figure}[h]
\centerline{\includegraphics{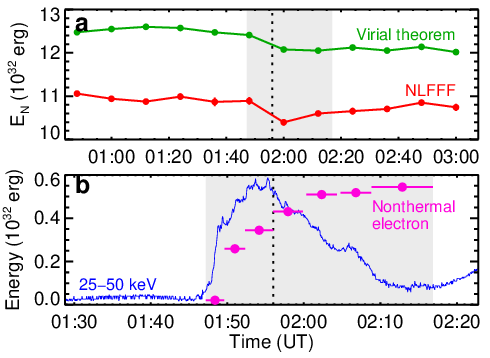}}
\caption{Estimation of flare energy release. (a) Magnetic energy estimated from the NLFFF extrapolation (same as Fig.~\ref{f:energy}) and the magnetic virial theorem. The magnetic virial theorem is applied on preprocessed lower boundary; the uncertainty is derived from a Monte-Carlo approach. Most error bars are too small to be seen. The grey background indicates the period during which the nonthermal electron energy is fitted in the next panel. (b) Accumulated nonthermal electron energy derived from \textit{RHESSI} HXR spectra using the thick-target model. Spectra for seven individual time intervals covering 01:44 UT to 02:17 UT (grey background) are fitted, the time range of each is noted by the horizontal bars. The HRX light curve in the 25--50 keV band is plotted in the background on a logarithmic scale. The vertical dotted line (01:56 UT) shows the SXR peak in GOES light curve.\label{f:budget}}
\end{figure}

To corroborate the result from the NLFFF extrapolation, we apply the magnetic virial theorem to the preprocessed lower boundary \citep[e.g.][and references therein; see also equation~\ref{eq:virial} in Appendix~\ref{a:definition}]{metcalf2005}. If the boundary is force-free enough and the region is largely isolated, as is the case for this HMI time series (see Appendix~\ref{a:extrapolation}), the free energy in the volume is expected to be well recovered \citep{metcalf2008}. Around the X-class flare, the evolution of $E_N$ computed from the virial theorem shows a similar trend to that from the NLFFF extrapolation, as shown in Figure~\ref{f:budget}a; the energy decreases during the flare are similar as well. $E_N$ derived from the virial theorem is about $14\%$ higher.

For the energetics during the flare, we independently estimate the nonthermal electron energy using the hard X-ray (HXR) spectra from the \textit{Ramaty High Energy Solar Spectroscopic Imager} \citep[\textit{RHESSI};][]{lin2002} under the classical thick-target model \citep[e.g.][and references therein]{holman2003}. The nonthermal electron flux spectra can be well fitted by a very steep power law with the index $\delta$$\sim$5--10 and a low energy cutoff $E_{\rm c}$$\sim$20--30 keV. The total nonthermal energy integrated from 01:44 UT to 02:17 UT is estimated to be $\sim$$0.54\times10^{32}\;\rm{erg}$, as shown in Figure~\ref{f:budget}b. Since the energy needed for thermal plasma heating, CME lift-off, and flare radiation are expected to be on the same order of magnitude, if not greater than the nonthermal electron energy \citep{emslie2004}, the estimated total energy budget has to be on the order of $10^{32}\;\rm{erg}$.

Based to this analysis, the computed pre-flare $E_f$ ($2.47\times10^{32}\;\rm{erg}$) is actually adequate to power this flare and accompanying CME eruption. On the other hand, the \textit{difference} between the pre- and post-flare $E_f$ ($0.34\times10^{32}\;\rm{erg}$), which we used as the estimate for energy loss, appears too small. One speculation is that the post-eruption coronal field is extremely dynamic and possibly deviates from the force-free condition, thus cannot be well described by a static force-free snapshot. If the field is locally (immediately above AR core) less energetic than what the NLFFF extrapolation indicates, then the discrepancy in the energy estimation will be alleviated. Another speculation involves the preprocessing scheme where a certain amount of smoothing is applied to the photospheric data to ensure the force-freeness of the boundary. Smoothing generally reduces the electric current inferred from the boundary, thus tending to lower the energy in the volume. Coordinated chromospheric magnetic field measurement from other observatories may help to constrain this procedure. A third possibility involves the photospheric field measurement itself. Small-scale field structures (sub-arc second) with high field gradients may carry strong electric current and a non-negligible amount of free energy, but will be unresolved with the moderate spatial resolution of HMI (1$\arcsec$), and thus overlooked in the modeling result.

We note that a series of comparative and diagnostic efforts pointed out the difficulties in employing realistic data input for NLFFF extrapolation \citep[e.g][]{schrijver2008nlfff,derosa2009}. Different algorithms can produce very different energy estimations and field topologies, which mainly arises from the inconsistency between the input boundary condition and the force-free assumption (Appendix~\ref{a:extrapolation}). We point out that the selection of the boundary data, e.g. the size of FOV, the degree of flux balance, can also directly affect the quality of the extrapolation result \citep{derosa2009}. Recent works by \cite{wiegelmann2010} and \cite{wheatland2011} included observational uncertainty in the revised algorithms. The former was tested with HMI data \citep{wiegelmann2011} and the later with SOLIS data \citep{gilchrist2011}. Both have shown to improve the quality/consistency of the solution. Further advancement of the algorithm is necessary and seems promising. Nevertheless, due to the aforementioned discrepancies, we think that quantitative energy estimations derived from modeling should be interpreted with care in the meanwhile.

\begin{figure*}
\centerline{\includegraphics{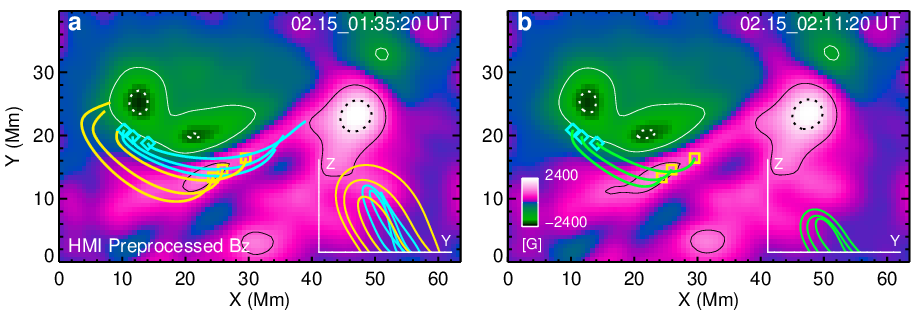}}
\caption{\textit{Schematic} illustration of the field connectivity as possible results from reconnection in the AR core region. (a) Selected NLFFF field lines prior to the flare, plotted on preprocessed vertical field. The contours are drawn at $\pm1000$ and $\pm2000\;\rm{G}$. The eastern set of loops (yellow) are expected to reconnect with the central set (cyan) and exchange footpoints. Side view of the loops from the left (east) is plotted as inset. (b) Selected post-flare loops (green) with footpoints that belonged to the two different sets of pre-flare loops in the previous panel. The loops become much shorter. These snapshots demonstrate that the reconnection scenario is consistent with the observed photospheric field. They are \textit{not} intended for identifying individual reconnecting loop pairs, nor modeling the reconnection process.
\label{f:reconnect}}
\end{figure*}

\subsection{On the Flare-Related Field Changes}
\label{subsec:implosion}

The rapid ($\sim$$10\;\rm{min}$) field change at the flare time likely has a coronal driver, as the photospheric Alfv\'{e}n timescale is too long to be compatible \citep[e.g.][]{hudson2011}. Reconnection, in particular, may change the coronal field connectivity fast enough to result in large-scale field variation at the ``line-tied'' loop footpoints. We illustrate one of the \textit{consequences} in Figure~\ref{f:reconnect} using the NLFFF extrapolation. Before the flare, highly sheared field lines (Fig.~\ref{f:reconnect}a) cross over each other near their footpoints, creating a configuration in favor of tether-cutting reconnection. Sigmoidal core loops ``are sheared past each other so that they overlap and are crossed low above the neutral line'' \citep{moore2001}. These loops exchange footpoints during the reconnection, and the inner footpoints indeed appear to be well linked in the post-flare field in the NLFFF extrapolation (Fig.~\ref{f:reconnect}b). New loops become shorter, consistent with the enhanced $B_h$ on the boundary. We note that the illustration is purely \textit{schematic} without any reference to the reconnection process itself. Moreover, loops on the north and south sides, further away from the PIL, may become connected after the flare, resulting in a more vertical configuration where $B_h$ decreases (Fig.~\ref{f:ribbon}a), similar to the effect of $\delta$-spot umbral darkening reported in \cite{liuc2005}. This scenario has been indirectly probed through the contraction of HXR sources \citep{ji2007} and the unshearing motion of EUV flare ribbons \citep{ji2006,su2007}. Direct imaging can be difficult as EUV or SXR images are usually saturated during major flares. The complicated temperature structures of the flaring atmosphere may also prevent a straightforward interpretation of the observations.

Besides the short loops near the PIL, tether-cutting reconnection also produces a flux-rope that connects the two far ends of the sigmoid after the exchange of the footpoints \citep{moore2001}. In a detailed account of the coronal activities in this event, \cite{schrijver2011x} use MHD modeling \citep{aulanier2010} to interpret the observations and find that the flux-rope formed by reconnection eventually erupts as part of the CME structure. Interesting evidence also arises from the ``sunquake'' observation reported by \cite{zharkov2011} following the initial report by \cite{kosovichev2011}. Two distinctive photospheric seismic sources are found at the two far ends of the sigmoid and their initiations closely match the flare/CME onset time. Because the signals appear before the HXR emission peak time and do not spatially match the HXR sources, they cannot be explained by particle precipitation as many earlier events are. Instead, flux-rope eruption appears to be the trigger, and magnetic variation is proposed as one of the possible mechanisms. The coronal fields rapidly reconstruct after the eruption and the overall effect may generate a strong downward impulse into the photosphere \citep{hudson2008}. Nevertheless, it is pointed out that the seismic sources do not appear to be co-spatial with the strongest magnetic field or the strongest field variations \citep{kosovichev2011}. The exact excitation mechanism remains to be explained. 

Regarding the possible ambiguity in the loop retraction observed in AIA 171 {\AA} (Fig.~\ref{f:contract}), we can rule out the scenario where loops remain steady but only cool sequentially into the EUV temperature sensitive passband. Such an explanation cannot account for the sudden start and stop of the motion, or the apparent loop oscillations. To rule out the projection effect, we inspect the coronal image sequences taken by SECCHI EUVI \citep{howard2008} on the twin \textit{STEREO} spacecraft that are both near quadrature from the Sun-Earth line. Coronal loops, when viewed from the side, appears to become ``flattened'' and in generally pressed down toward the solar surface during the eruption. The apparent loop length, if unchanged, should increase instead. Therefore the retraction is likely to be real, its amplitude and speed underestimated from overhead AIA observation. Detailed analysis of EUV loop motion may provide quantitative diagnostics of the coronal field \citep[e.g][]{aschwanden2011}, but is out of the scope of this work.

We note that the field change discussed here must be distinguished from the transient variations in the longitudinal field \citep{kosovichev2011}. The latter lasts only a couple of tens of minutes and is likely an artifact caused by flare-related emissions. HMI spectra at these abnormal pixels briefly deviate from the usual absorption profile, showing an enhanced line center as expected \citep{maurya2011}.


\section{Summary}
\label{subsec:summary}

We have studied the magnetic field and its energy of AR 11158 over a period of 5 days using a series of HMI vector magnetograms. A NLFFF extrapolation, coupled with coronal imaging, provides information of the coronal magnetic structure, electric currents and free energy. From its early flux emergence to recurrent major flares and CME eruptions, the AR displays distinctive stages of energy build-up and release, driven by or resulting in gradual or sudden magnetic field changes. We summarize the major results as follows:

\begin{itemize}
\renewcommand{\labelitemi}{$-$}
\item The quadrupolar AR primarily consists of two interacting bipoles. The trailing polarity of the leading pair and the leading polarity of the trailing pair undergo significant shearing. A pronounced filament appears over the major PIL early and its main part persists through multiple eruptions, including the X-2.2 class flare and halo CME on 2011 February 15. NLFFF extrapolation suggests that the field in the filament channel carries strong current, and is highly sheared with about 0.9 turn near the axis.

\item NLFFF extrapolation indicates significant electric current and free energy injection during early flux emergence, about $10^{32}\;\rm{erg}$ over a mere 10 h. Current and energy mostly concentrate in or near the filament channel in the low atmosphere: over $50\%$ is stored below $6\;\rm{Mm}$. The computed peak free energy reaches about $2.59\times10^{32}\;\rm{erg}$. A $0.34\times10^{32}\;\rm{erg}$ decrease is found within 1 hour after the X-class flare. We show that the effect of random noise is small, but the systematic uncertainties can be large. The energy loss is most likely underestimated.

\item Rapid and irreversible enhancement of the horizontal field takes place along the PIL during the X-class flare. The increase in $B_h$ is $28\%$ on average; only a $5\%$ decrease is found for $|B_z|$ and the signal is mixed. The observed photospheric field becomes overall stronger, more inclined and better aligned with the PIL. The short time scale indicates a coronal driver. Shorter loops created by tether-cutting type reconnection are \textit{a priori} consistent with the photospheric signatures. The reconnection picture is supported by the fact that the change largely happens in the area that are either swept by the flare ribbons or lying in between.

\item NLFFF extrapolation demonstrates that the change in the photospheric and coronal field is morphologically consistent with the ``magnetic implosion'' conjecture. Energy loss during the explosions cause the coronal field to contract to reach a new balance. This scenario is supported by the EUV loop retractions observed from AIA. The extrapolated field appears to be more ``compact'' after the flare, the lowest layer is more sheared but it relaxes faster with height and is overall less energetic.
\end{itemize}

Regular quantitative study and statistical surveys of a large ensemble of AR vector magnetic fields now become possible with HMI observing the full solar disk at high spatial and temporal resolution. Although problems still abound in both data reduction and modeling procedures, we anticipate more in-depth studies using the newly available data will eventually lead to better understanding of the AR magnetic fields and energetics. 


\acknowledgments

Data and images are courtesy of NASA/SDO and the HMI and AIA science teams. We thank the anonymous referee for the helpful comments. We thank J. Schou, S. Couvidat and R. C. Elliot for the help on HMI noise characterization and Monte-Carlo experiment, W. Liu for the help on AIA and SOT data. \textit{SDO}/HMI is a joint effort of many teams and individuals to whom we are greatly indebted. \textit{Hinode} is a Japanese mission developed and launched by ISAS/JAXA, with NAOJ as domestic partner and NASA and STFC (UK) as international partners. It is operated by these agencies in co-operation with ESA and NSC (Norway). \textit{RHESSI} is a NASA Small Explorer (SMEX) mission. This work is supported by NASA Contract NAS5-02139 (HMI) to Stanford University.

{\it Facilities:} \facility{HMI (\textit{SDO})}, \facility{AIA (\textit{SDO})}.



\appendix

\section{Evaluation and Discussion of the Extrapolation Algorithm}
\label{a:extrapolation}

Previous studies of coronal field extrapolation have shown the importance of appropriate input boundary conditions \citep[e.g.][]{metcalf2008,derosa2009}. It is suggested that the input magnetograms should: \romannumeral 1) have a large FOV with sufficient information on the horizontal component even in the weak field region; \romannumeral 2) have balanced magnetic flux; and \romannumeral 3) have balanced Lorentz force and torque. The HMI vector magnetograms used in this study appear to be a good candidate. For the February 14 20:35 UT frame, the region of interest with strong field ($|B_z|>100\;\rm{G}$) covers the center $11\%$ area of the FOV (a total of $216\times216\;\rm{Mm^2}$) and is isolated from the side boundaries. There are no extended plage regions or ARs nearby, and the large outskirts included correspond to a coronal volume that contains most of the observed EUV loops. The net magnetic flux accounts for only $0.3\%$ of the total unsigned flux for the strong field region, thus the flux is well balanced. In particular, the surface integrated Lorentz force and torque are 5--15 times lower than the several data sets used in earlier studies (from different instruments, for different ARs) \citep{wiegelmann2011}. We note that such low Lorentz force/torque may be a special case, as the photospheric field is generally not expected to be force-free \citep{metcalf1995}. For the preprocessing scheme \citep{wiegelmann2006} applied before the extrapolation, we test and adopt the following parameter set: $\mu_1=\mu_2=1$, $\mu_3=0.001$, and $\mu_4=0.01$. Here $\mu_1$ and $\mu_2$ control the net force and torque on the boundary respectively; $\mu_3$ controls the relative influence of the observed data, and $\mu_4$ the level of smoothing.

We evaluate the quality of the extrapolation by computing three metrics: the mean Lorentz force $L_f$, the mean field divergence $L_d$, and the current weighted mean angle between the magnetic field and electric current $\sigma_j$:
\begin{equation}
\label{eq:lf}
L_f=\left \langle \frac{|\bm{B} \times (\nabla \times \bm{B})|^2}{B^2} \right \rangle,
\end{equation}
\begin{equation}
\label{eq:ld}
L_d=\left \langle |\nabla \cdot \bm{B}|^2 \right \rangle,
\end{equation}
\begin{equation}
\label{eq:sigmaj}
\sigma_j=\sin^{-1} \left ( \left \langle \frac{|\bm{J}\times\bm{B}|}{B} \right \rangle / \langle J \rangle \right ),
\end{equation}
where $\bm{B}$ is the NLFFF field vector, $B=|\bm{B}|$, $J=|\bm{J}|$, and angle brackets denote the mean value within the domain. Following literature, we normalize $L_f$, $L_d$ to a unit volume and normalize $L_d$ further by $1\;\rm{G^2}$. Ideally all should vanish. For a total of 120 hourly frames computed in this study, we find $L_f=6.4\pm2.2$, $L_d=3.1\pm1.2$, and $\sigma_j=18.0^\circ\pm1.2^\circ$. These values are typical for this optimization-based algorithm and are comparable to previously reported results based on other solar data sets \citep[e.g.][]{schrijver2008nlfff}. 

It has been pointed out that even after preprocessing, the observed field, with its uncertainties, is still incompatible with a strictly force-free field \citep{derosa2009}. The fact that a better suited input does not lead to noticeable improvement in this case may suggest an internal limit to the current algorithms when dealing with realistic, imperfect boundary conditions. In light of this, a new implementation of the optimization scheme \citep{wiegelmann2010} has been developed and tested for HMI data. After including information on the horizontal field measurement uncertainties, the solution proves to satisfy the force-free condition significantly better \citep{wiegelmann2011}. For the 20:35 UT frame,  $\sigma_j$ is reduced from $16.6^\circ$ to $5.7^\circ$.

Problems may also arise when we adopt a large computation domain but still assume a planar boundary. A magnetogram spanning about $18^\circ$ in Heliographic longitude, as used in this study, will then have its edge elevated by $(\sec{9^\circ}-1)R_\odot$ ($8.7\;\rm{Mm}$, or 12 pixels) above the solar surface. This may be a reason that we fail to faithfully reproduce the long loops on the north side. These loops extend far away from the AR center and appear to be low-lying judged from the \textit{STEREO}/EUVI observations. They are thus more subject to inconsistencies in the boundary condition. We note that the free energy estimation may be less effected in this case, as these loops are largely potential-like. Nevertheless, as multiple ARs are often magnetically connected in a global scale, implementation of the model in spherical geometry \citep[e.g.][]{tadesse2011} becomes necessary.

\section{Computation of Various Magnetic Field Properties}
\label{a:definition}

The torsion parameter, $\alpha$, indicates the twist and non-potentiality of the force-free magnetic structure. It may be computed from any component of Equation~\ref{eq:nlfff}. For its distribution on the photosphere, we use the vertical ($z$) component. For the filament cross section shown in Figure~\ref{f:filament}e, assuming the normal vector of the vertical plane is $\bf{n}$ (which itself is horizontal), we use the horizontal component:
\begin{equation}
\label{eq:alpha}
\alpha=\frac{(\nabla\times\bm{B})_h \cdot \bf{n}}{\bm{B}_h \cdot \bf{n}}.
\end{equation}

The $\tilde{N}$ metric \citep{demoulin1996} is used to quantify the spatial gradient in magnetic connectivity. When the mapping of magnetic field line diverges, $\tilde{N}$ increases drastically and indicates the existence of a quasi-separatrix layer (QSL). For a footpoint $P(x,y,0)$ on the bottom boundary in the domain, we integrate the field line passing through it and obtain the other footpoints $(x^\prime,y^\prime,0)$. Field lines passing through the side walls or upper boundary are treated as missing values in this work. With the displacement vector defined as $\{X_1,X_2\}=\{x^\prime-x,y^\prime-y\}$, $\tilde{N}$ is defined as:
\begin{equation}
\label{eq:norm}
\tilde{N}(x,y,0)=\sqrt{\sum\limits_{i=1,2}\left[\left(\frac{\partial X_i}{\partial x}\right)^2+\left(\frac{\partial X_i}{\partial y}\right)^2\right]}.
\end{equation}

The shear angle measures the angular deviation between a field vector from the potential counterpart. Let $\bm{B}_{N}$ and $\bm{B}_{P}$ denote the NLFFF and PF vector, $|\bm{B}_{N}|=B_N$, $|\bm{B}_{P}|=B_P$, the mean shear angle $\theta$ at each point is defined as:
\begin{equation}
\label{eq:shear}
\theta=\cos^{-1}\displaystyle \frac{\bm{B}_{N} \cdot \bm{B}_{P}}{B_N\,B_P}.
\end{equation}
The weighted mean shear $\theta_w$ is computed by summing over all pixels in the domain \citep{wanghm1994}:
 \begin{equation}
\label{eq:wshear}
\theta_w=\frac{\sum {\theta\,B_N}} {\sum {B_N}}.
\end{equation}

The magnetic virial theorem can be used to calculate the force-free field energy in a volume solely from its boundary. When applied to the lower boundary ($z=z_0$) in a Cartesian coordinate, it can be expressed as
\begin{equation}
\label{eq:virial}
E=\frac{1}{4 \pi} \int_{z=z_0} (xB_x+yB_y)\,B_z\,{\rm d}x\,{\rm d}y,
\end{equation}
where $E$ is the force-free field energy, and $x$, $y$ are the coordinate of where the field is measured. Following \cite{metcalf2005}, we estimate the uncertainty in $E$ through a pseudo-Monte Carlo method. We randomly vary the origin point of the coordinate system and repeat the evaluation for large numbers. The standard deviation is adopted. It should be 0 if the field is perfectly force-free, and no magnetic flux connects outside the domain.









\end{document}